\documentclass[aps,epsfig,float,twocolumn]{revtex4}

\usepackage{graphicx}

\newcommand{\ltsims}{\protect\raisebox{-0.5ex}{$\stackrel{\scriptstyle <}{\sim}$}}
\newcommand{\gtsims}{\protect\raisebox{-0.5ex}{$\stackrel{\scriptstyle >}{\sim}$}}

\def\i{\mathrm{i}}
\def\e{\mathrm{e}}

\def\s{\sigma}
\def\exp#1{\mathrm{exp}\left\{ #1 \right\}}
\def\dsum#1#2{\displaystyle\sum_{#1}^{#2}}
\def\dfrac#1#2{\displaystyle\frac{#1}{#2}}
\def\[{\left[}
\def\]{\right]}

\begin{document}
\title{Pressure-induced phase transition and bi-polaronic sliding in a hole-doped Cu$_2$O$_3$ ladder system}
\author{E. Kaneshita, I. Martin, and A. R. Bishop}
\affiliation{
Los Alamos National Laboratory, Los Alamos, NM 87545}
\date{\today}
\begin{abstract}
We study a hole-doped two-leg ladder system including metal ions, oxygen, and electron-lattice interaction, as a model for Sr$_{14-x}$Ca$_x$Cu$_{24}$O$_{41-\delta}$.
Single- and bi-polaronic states at $\frac{1}{4}$-hole doping are modeled as functions of pressure by applying an unrestricted Hartree-Fock approximation to a multiband Peierls-Hubbard Hamiltonian.
We find evidence for a pressure-induced phase transition between single-polaron and bi-polaron lattices.
The electronic and phononic excitations in those states, including distinctive local lattice vibrational modes, are calculated by means of a direct-space Random Phase approximation.
Finally, as a function of pressure, we identify a transition between site- and bond-centered bi-polarons, accompanied by a soft mode and a low-energy charge-sliding mode.
We suggest comparisons with available experimented data.
\end{abstract}
\maketitle

\section{introduction}
A material of considerable recent interest is the compound Sr$_{14-x}$Ca$_x$Cu$_{24}$O$_{41}$ (SCCO), which shows superconductivity under pressure~\cite{uehara,nagata}.
The SCCO structure includes quasi-one dimensional (Q1D) two-leg ladders Cu$_2$O$_3$ and one-dimensional (1D) CuO$_2$ chains, while other high-$T_c$ cuprate superconductors found so far contain two-dimensional (2D) CuO$_2$ planes.
The Cu$_2$O$_3$ ladders and CuO$_2$ chains in this system are intrinsically hole-doped even at $x=0$ with a total of 6 holes per formula unit.
(The total hole concentration can be decreased by La and Y substitution for Sr, with \textit{e.g.} La$_{6}$Cu$_{8}$O$_{41}$ containing no holes.)
Of these, only approximately one hole goes into the ladder component of the formula unit, (Cu$_2$O$_3$)$_7$, which results in an effective doping of about 7\% per Cu site in the ladder~\cite{nucker,osafune}.
Ca substitution, $x$, does not change the total number of carriers, but transfers the holes from the chains to the ladders~\cite{osafune,mizuno,ruzicka}.
The conductivity increases with increasing $x$~\cite{yamane}.

The doped holes can create a polaronic or charge-density-wave (CDW) state, and a charge sliding mode could be expected as a collective excitation.
The existence of such states and modes is supported by some experiments in the material with $x=0$.
Resonant X-ray scattering has revealed a five-site periodic hole structure in the ladder~\cite{abbamonte,rusydi}.
Microwave measurements show a relatively small $c$-axis conductivity with a narrow peak in a very low-energy region ($\sim$ 0.2 meV)~\cite{kitano1,kitano2}.
This low-energy resonance is observed up to a temperature ($\sim$ 100 K) too high to be attributed to single particle excitations, which would be completely broadened and no longer observed above 10K ($\sim$1meV) due to thermal fluctuations.
Similarity has been noted between the nonlinear behavior of the conductivity in SCCO at $x=0$ and that of the sliding mode in materials supporting CDW states.
Blumberg \textit{et al.} have reported~\cite{blumberg} that the low-frequency dielectric constant ($\epsilon_0 \sim 4 \times 10^{6}$) obtained by transport measurements is consistent with estimates from the pinning energy ($\sim$ 0.2 meV) suggested by microwave measurements.

There are several possible polaronic configurations.
One is to randomly distribute the polarons.
Another is that some of them bind to compose various multi-polaron configurations:
In a 2D case, the doped holes can arrange into stripes, which is one kind of multi-polaron state.
The groundstate configuration depends on the doping level, pressure, \textit{etc}.
Modeling the pressure-dependence of the groundstate configuration is one aim of this study.

In a previous work~\cite{martin,kane}, we studied the electronic and phononic excitations in a 2D CuO$_2$ plane with inhomogeneous charge-lattice-spin structures (stripes and other polaron patterns).
We identified local ``edge" or ``interface" modes in the phononic and electronic (spin and charge) excitations induced by the inhomogeneity.
In the Cu$_2$O$_3$ two-leg ladder in SCCO, we can similarly anticipate the existence of inhomogeneous structure and associated local excitation modes in spin, charge, and lattice degrees of freedom.

Here, we investigate the ground states and the electronic and phononic excitations in the two-leg ladder system by applying an unrestricted Hartree-Fock and a direct-space Random Phase approximations (RPA) to a multi-band Peierls-Hubbard Hamiltonian.
We consider single-polaron (SP) and bi-polaron (BP) states:
The latter comprise bound polarons extending over rungs.
Compared to the 2D cases, the SP state that includes isolated polarons, is found to possess similar phonon excitations as the diagonal stripe state or the periodic polaron state~\cite{martin,kane}.
On the other hand, the BP state shows the same type of local phonon modes as the vertical stripe state in the 2D system.
This is reasonable, since the vertical stripe state is a form of multi-polaron state, which includes several polarons coupled by shared O ions.
To model the effect of pressure on the groundstate configuration, we compare the energies of the SP state and the BP state, while varying the Cu-O hopping integral.
We find that as a function of increasing ``pressure" (modeled by increasing hopping strength) a transition from the SP state to the BP state is induced, together with interesting intermediate states.
Most strikingly, we identify a transition between site- and bond-centered BPs accompanied by phonon softening indicative of the onset of sliding or other instabilities.

\section{formulation}
\subsection{Hamiltonian}
To study a Cu$_2$O$_3$ ladder, we use the following three-band extended Peierls-Hubbard Hamiltonian, which
includes both electron-electron and electron-lattice
interactions~\cite{yone,yi}:
\begin{eqnarray}\label{eq:H0}
H_0 = \sum_{<ij> \sigma} t_{pd}(u_{ij})
(c^\dagger_{i\sigma}c_{j\sigma}+ H.c.)
+\sum_{i,\sigma} \epsilon_i(u_{ij})
c^\dagger_{i\sigma}c_{i\sigma} \nonumber\\
+\sum_{<ij>} \frac{1}{2} K_{ij}u^2_{ij}
+\sum_{i, j,\sigma, \sigma'}{U_{ij}n_{i\sigma}n_{j\sigma'}}.
\end{eqnarray}
We impose periodic (open) boundary condition in the $x$ ($y$) direction, \textit{i.e.}, there are two periodic Cu-O chains along the $x$-direction (we term this oxygen O$_x$ subsequently), connected together through the other oxygens (O$_y$).
In this Hamiltonian, $c^\dagger_{i\sigma}$ creates a hole with spin $\sigma$ on site $i$, and each site has one orbital (d$_{x^2-y^2}$ on Cu, and p$_x$ or p$_y$ on O). The Cu (O) site electronic energy is $\epsilon_d$ ($\epsilon_p$). $U_{ij}$ represents the on-site Cu (O) Coulomb, $U_d$ ($U_p$), or the nearest-neighbor repulsion, $U_{pd}$. The electron-lattice interaction modifies the Cu-O hopping strength linearly through the oxygen displacement $u_{ij}$: $t_{pd}(u_{ij}) = t_{pd}(1\pm \alpha u_{ij})$, where $+(-)$ applies if the Cu-O bond shrinks (stretches) for a positive $u_{ij}$; it also affects the Cu on-site energies $\epsilon_d(u_{ij}) = \epsilon_d + \beta\sum_j{(\pm u_{ij})}$, where the sum is over the three neighboring O ions.
Other oxygen modes (buckling, bending, etc) are assumed to couple to electron charge more weakly and are neglected here for simplicity, but can be included as necessary within the same approach.
We use variations around the following set of model parameters used in 2D CuO$_2$ models~\cite{martin,kane}: $\epsilon_p-\epsilon_d = 4$ eV, $U_d =8$ eV, $U_p = 3$ eV, $U_{pd} = 1$ eV, and $K = 32$ eV/{\AA$^2$}, $\alpha = 2.0$ eV/\AA, $\beta = 1$ eV/\AA; we vary $t_{pd}=1\sim5$eV to simulate the pressure effect.
This is clearly an oversimplified representation of pressure effects, but serves to demonstrate the ground state phases and transitions.
Effects of varying the coupling strength are also considered below with similar results.
To approximately solve the above model, we use unrestricted Hartree-Fock combined with an inhomogeneous generalized RPA to study linear fluctuations of lattice, spin or charge~\cite{yone} in a supercell of size $N_x \times 2$ (we take $N_x=8$ here).

\subsection{Phonon spectral function}
The output of the calculation is the Hartree-Fock ground state and the linearized fluctuation eigen-frequencies and eigen-vectors with respect to it.
From the \textit{phonon} eigen-modes, we calculate the corresponding neutron
scattering cross section:
\begin{eqnarray}
S(\mathbf{k},\omega) =\int dt\, e^{-i\omega t}
\sum_{l l^\prime}{\langle e^{-i \mathbf{k r}^{\epsilon}_l(0)}
e^{i\mathbf{k r}^{\epsilon}_{l'}(t)}\rangle},
\end{eqnarray}
where $\epsilon$ labels the five ions in the unit cell of the ladder: (1) O$_x$ ions in the lower leg, (2) O$_y$ ions in rungs, (3) O$_x$ ions in the upper leg, (4) Cu ions in the lower leg, and (5) Cu ions in the upper leg;
Here the position of each ion is expressed by $\mathbf{r}^{\epsilon}_{l}(t) = \mathbf{x}_{l} + \mathbf{d}^{\epsilon} + \mathbf{u}^{\epsilon}_l(t)$, where each of the terms represents the location of the $l$-th unit cell origin $\mathbf{x}_{l}$($=x_l\hat{1}_x$), time-dependent vibrational component $\mathbf{u}^{\epsilon}_l(t)$, and position within the unit cell $\mathbf{d}^{\epsilon}$:
\begin{eqnarray}
&&\mathbf{d}^{(1)}= \frac{a}{2} \hat{1}_x,\,\,\,\, \mathbf{d}^{(2)}=
\frac{a}{2}\hat{1}_y ,\,\,\,\,
\mathbf{d}^{(3)}= \frac{a}{2} \hat{1}_x+ a \hat{1}_y,\nonumber\\
&&\hspace{1cm}\mathbf{d}^{(4)}=\hat{0},\,\,\,\,
\mathbf{d}^{(5)}=a\hat{1}_y.
\end{eqnarray}

As noted above, for simplicity we consider Cu ions as fixed, and the motion of O ions oriented along the corresponding Cu-O bond:
$\mathbf{u}^{\epsilon}_l = u^{\epsilon}_l \hat{\e}_{\epsilon}$ with
$\hat{\e}_1=\hat{\e}_3=\hat{1}_x$, $\hat{\e}_2=\hat{1}_y$,
$\hat{\e}_4=\hat{\e}_5 =\hat{0}$. The scalar displacements can now be expressed in terms of the normal modes $z_n$ as $u^{\epsilon}_l(t) = \sum_n{\Phi^{\epsilon}_{x_l,n}z_n(t)}$. Making a second-order expansion in the oxygen displacements, we obtain
\begin{eqnarray}
&&\hspace{-0.5cm}S(\mathbf{k},\omega)=\nonumber\\
&&\sum_{n}\Big\{
\Big[\,k_x^2\,|\Phi^{(1)}_{k_x,n}|^2
+k_y^2\,|\Phi^{(2)}_{k_x,n}|^2
+k_x^2|\Phi^{(3)}_{k_x,n}|^2
\Big]\nonumber\\
&&\hspace{0.75cm}+\Big[
k_xk_y
\,\e^{-\i (k_x-k_y) \frac{a}{2}}
\,\Phi^{(1)}_{k_x,n} (\Phi^{(2)}_{k_x,n})^*
+c.c.\Big]\nonumber\\
&&\hspace{1cm}
+\Big[k_xk_y
\,\e^{\i (k_x\frac{a}{2}+ k_y \frac{a}{2})}
\,\Phi^{(2)}_{k_x,n} (\Phi^{(3)}_{k_x,n})^*
+c.c.\Big]\nonumber\\
&&\hspace{1.25cm}
+\Big[k_x^2
\,\e^{-\i k_y a}
\,\Phi^{(3)}_{k_x,n} (\Phi^{(1)}_{k_x,n})^*
+ c.c.
\Big]
\Big\}\nonumber\\
&&\hspace{0.5cm}\times\frac{\hbar}{2m\omega_n}
[(1+n_B)\delta(\omega-\omega_n) + n_B\delta(\omega+\omega_n)].
\end{eqnarray}
Here, $\Phi^{\epsilon}_{k_x,n}  = \sum_l{e^{-\i k_x
x_l}\Phi^{\epsilon}_{x_l,n}}$, and $n_B = (e^{\omega_n/T}-1)^{-1}$ is the
thermal population of the phonon mode $n$. This is a generalization of the
usual neutron scattering intensity expression~\cite{lovesay} for the case of
phonons with a larger real space unit cell. We plot
$S(\mathbf{k},\omega)/|\mathbf{k}|^2$ for \textbf{k}-directions sampling
longitudinal modes, consistent with the common experimental convention.

\subsection{Electron spectral function}
To investigate the neutral \textit{electronic} excitations, we calculate the spectral function~\cite{fetter}:
\begin{eqnarray}
\sum_n |\, \langle \Psi_0| \mathcal{O}(\mathbf{k})|\Psi_n \rangle \,|^2\,\delta(\omega-E_n-E_0),
\end{eqnarray}
where $|\Psi_0\rangle$ ($|\Psi_n\rangle$) is the RPA ground (excited) state whose energy is represented by $E_0$ ($E_n$), and $\mathcal{O}(\mathbf{k})$ is an operator, e.g. spin $\mathbf{S}(\mathbf{k})$ or charge $n(\mathbf{k})$, summed over Cu- and O-sites:
\begin{eqnarray}
\mathcal{O}(\mathbf{k})=\sum_{\epsilon\,=\,1}^5
\mathcal{O}^{\epsilon}(\mathbf{k})\,e^{-\i \mathbf{kd}^\epsilon}.
\end{eqnarray}
The effect of an infinitesimal external field corresponding to the excitation $\Psi_n$ can be represented by the change of an observable $\langle\mathcal{O}\rangle$ in the state $\Psi=\Psi_0 + \eta \Psi_n$ ($|\eta|\ll1$):
\begin{eqnarray}
\langle\mathcal{O}\rangle\simeq\langle\mathcal{O}\rangle_0+\delta\langle\mathcal{O}\rangle_n,\\
\delta\langle\mathcal{O}\rangle_n\propto\langle\Psi_0| \mathcal{O} |\Psi_n\rangle,
\end{eqnarray}
where $\langle\mathcal{O}\rangle_0$ is the expectation value with respect to the ground state.

As noted earlier, we simulate the effect of pressure by varying $t_{pd}$.
We identify a transition from the SP to the BP state with distinct electronic and phononic signatures.
Note that the BP states here are polaron bound states on different legs of the ladder (in contrast to same-chain BPs).

\section{results}
\subsection{Polaronic ground states and phase transitions}
\begin{figure}[htb]
\begin{center}
\includegraphics[width = 0.7\linewidth]{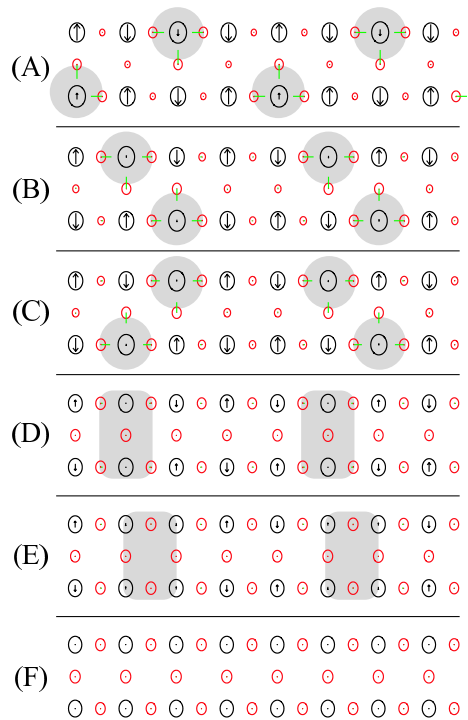}
\caption{(Color online) ground states for several values of $t_{pd}$:
(A) $t_{pd}=1.2$, (B) $t_{pd}=1.6$, (C) $t_{pd}=1.8$, (D) $t_{pd}=3.0$, (E) $t_{pd}=3.5$, and (F) $t_{pd}=4.7$.
The circles (radius) and the arrows (length and direction) represent the hole and spin densities, respectively.
Lines at the positions of O-ions represent static displacements, and shadows show where the polarons are located.
}\label{fig:pht-gs}
\end{center}
\end{figure}
We first show in Fig.~\ref{fig:pht-gs} the configurations of the ground states obtained by the Hartree-Fock calculation for several values of $t_{pd}$.
For $1.0<t_{pd}<5.0$, there are six types of groundstate configurations.
(A) in Fig.~\ref{fig:pht-gs} is a single-polaron (SP) state, which has $n$ polarons in a staggered arrangement for an $n$-hole doped system ($n=4$ here).
(B) and (C) are diagonal bi-polaron (DBP) states, which have diagonally-bound polarons; and (D) and (E) are vertical bi-polaron (VBP) states, which have the same structure as a short segment of a vertical stripe in the 2D system.
(D) consists of site-centered VBPs, and (E) of bond-centered VBPs.
(F) is the uniform (UNI) state, which does not have any local spin or charge modulation, or lattice displacement.

Note that the undoped system shows the AF configuration without any lattice displacement (not shown here).
One can expect that a singlet solid is the more likely exact undoped ground state, however the Hartree-Fock calculation favors an AF.

\begin{figure}[htb]
\begin{center}
\includegraphics[width = 0.9\linewidth]{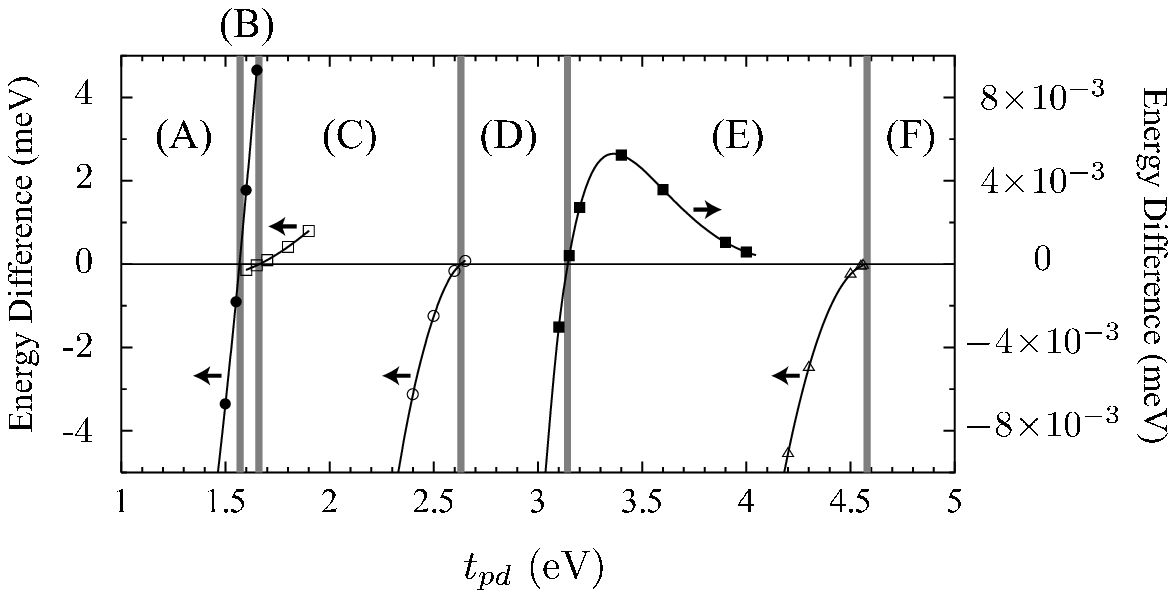}
\caption{$t_{pd}$-dependence of the energy difference between the (meta-) stable states with $\alpha=2.0$: $E_{\mathrm{(A)}}-E_{\mathrm{(B)}}$ (filled circles), $E_{\mathrm{(B)}}-E_{\mathrm{(C)}}$ (open squares), $E_{\mathrm{(C)}}-E_{\mathrm{(D)}}$ (open circles), $E_{\mathrm{(D)}}-E_{\mathrm{(E)}}$ (filled squares), and  $E_{\mathrm{(E)}}-E_{\mathrm{(F)}}$ (open triangles).
$E_{\mathrm{(D)}}-E_{\mathrm{(E)}}$ is of order $10^{-3}$ meV per unit cell (Cu$_2$O$_3$), while the others are of order $10^{-1}\sim1$ meV.
Each state (A)-(F) is the ground state in the corresponding region separated by the thick grey lines.
The lines are guides to the eye only.
}\label{fig:pht-e}
\end{center}
\end{figure}
With $E_{(\mathrm{A})}$, $E_{(\mathrm{B})}$, ..., $E_{(\mathrm{F})}$ the energies of (A), (B), ..., (F), respectively, we compare these energies, and determine the regions of $t_{pd}$ corresponding to the (A)-(F) phases for $1.0<t_{pd}<5.0$.
The $t_{pd}$-dependence of these groundstate configurations is shown in Fig.~\ref{fig:pht-e}.

There are five groundstate transitions for $1.0<t_{pd}<5.0$.
These are first-order transitions except for the one between (E) and (F),which is of second order.
The main feature here is that the larger $t_{pd}$, the more delocalized the ground state becomes.
This follows from the fact that the transition with increasing $t_{pd}$ is SP $\rightarrow$ BP $\rightarrow$ UNI.
By studying the polaron eigen-functions, we find that the transitions occur when the SP-SP or BP-BP overlap achieves sufficient levels, resulting in the polaron melting ,\textit{i.e.}, the sequence of the transitions SP $\rightarrow$ BP $\rightarrow$ UNI.

The transition DBP $\rightarrow$ VBP with increasing $t_{pd}$ is similar to the transition from diagonal to vertical stripe with increasing hole-concentration in some of the high-$T_c$ cuprates with CuO$_2$ planes~\cite{machida,matsuda}.
The energy difference between (D) and (E) near the transition point is smaller than for the other transitions.
$E_{(\mathrm{D})}-E_{(\mathrm{E})}$ and $E_{(\mathrm{E})}-E_{(\mathrm{F})}$ both asymptotically vanish, as $t_{pd}$ approaches the (E)-(F) transition point from smaller $t_{pd}$.
This transition is associated with recovery of the broken symmetry.
Below, we show the existence of the VBP sliding mode, which recovers the translational symmetry.

\subsection{Sliding mode in bi-polaronic states}
\begin{figure}[b]
\begin{center}
\includegraphics[width = 0.8\linewidth]{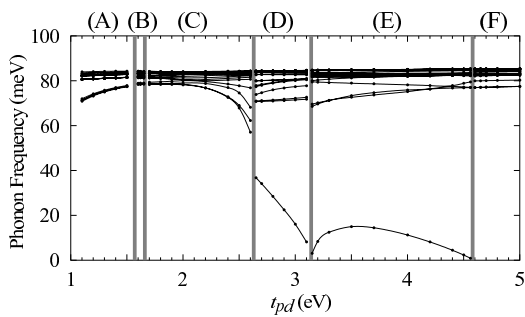}
\caption{$t_{pd}$-dependence of the phonon eigen-frequencies.
The lines are guides to the eye only.
}\label{fig:pht-w}
\end{center}
\end{figure}
\begin{figure}[htbp]
\begin{center}
\includegraphics[width = 0.8\linewidth]{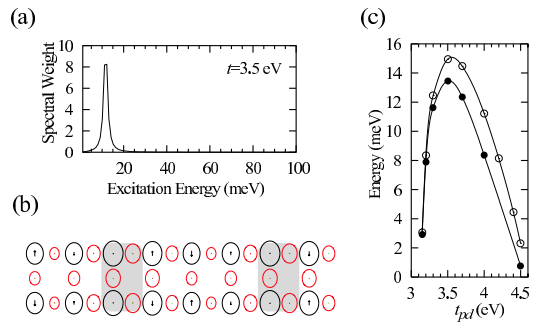}
\caption{(Color online) (a) charge excitation for $\mathbf{k}=(\frac{\pi}{2},0)$ at $t_{pd}=3.5$ eV from the RPA calculation.
(b) Excited state at $E\sim13$ meV.
The ground state is (E) in Fig.~\ref{fig:pht-gs}.
Comparing this excited state with the ground state, a sliding of BPs is found.
(c) Energies of soft phonon modes (open circles) and charge sliding modes (filled circles).
}\label{fig:sliding}
\end{center}
\end{figure}

The $t_{pd}$-dependence of the phonon eigen-frequencies obtained by a direct-space RPA calculation is shown in Fig.~\ref{fig:pht-w}.
A main branch lies in the range of 80-85 meV and is consistent with the main phonon branch observed experimentally.
This frequency range is insensitive to the doping level.
However, some characteristic \textit{local} modes are induced below this frequency range by hole doping, similar to the 2D cases~\cite{martin,kane}.
Additionally, an extremely soft mode is found in the (D) and (E) phases.
The frequency of the soft phonon mode in the (E) configuration is purely imaginary in the (D) phase region, and vice versa.
To further understand this soft mode, we calculate the corresponding electronic excitations [Fig.~\ref{fig:sliding}].
We find one soft charge excitation [Fig.~\ref{fig:sliding} (a)], which shows a sliding mode [Fig.~\ref{fig:sliding} (b)] and whose excitation energy shows the same behavior as that of the soft phonon mode.
We identify the soft phonon mode as one coupling with a sliding mode of VBPs along the ladder.
We identify the frequency of this mode as a pinning energy of the VBP sliding, corresponding to a Peierls-Nabarro barrier from the lattice discreteness.
In the (D) phase [(E) phase], VBPs are pinned Cu site-centered [bond-centered] by a potential energy; the potential energy is minimal at a Cu site [an O$_x$ site] and maximal at an O$_x$ site [a Cu site].
The minimum and maximum points of the potential energy exchange at the (D)-(E) transition point.
We will discuss this further below.

From the softening of the sliding mode and the second-order-like behavior of the energy differences, $E_{(\mathrm{D})}-E_{(\mathrm{E})}$ and $E_{(\mathrm{E})}-E_{(\mathrm{F})}$ near the (E)-(F) transition point, we conclude that the pinning potential becomes flat at that transition point.

\subsection{Effects of electron-lattice coupling}
We next consider effects of the electron-lattice coupling on the groundstate configuration.
For this purpose, changing the coupling strength $\alpha$ (=1.0, 2.0, 3.0, 4.5) we calculate the critical values of $t_{pd}$ in the same manner as above.
In this way, the $t_{pd}-\alpha$ phase diagram is found as in Fig.~\ref{fig:pht-gs}.
Increasing the coupling strength tends to raise the critical values of $t_{pd}$ except for that between (E) and (F):
The boundary between (E) and (F) is insensitive to the change of the coupling strength.
The difference in the $\alpha$-dependence shows that the transition between (E) and (F) has a different character than other transitions:
As seen above, the transition between (E) and (F) is most likely of second-order.
\begin{figure}[h]
\begin{center}
\includegraphics[width = 0.7\linewidth]{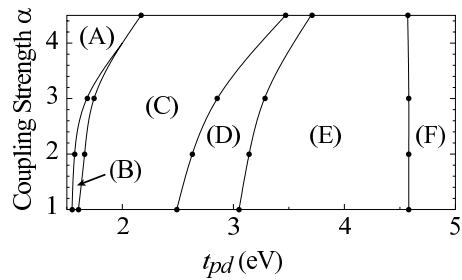}
\caption{phase diagram obtained from the calculation of the critical values of $t_{pd}$ for various electron-lattice coupling strength $\alpha$.
(A)-(F) correspond to the states in Fig.~\ref{fig:pht-gs}.
The lines are guides to the eye only.
}\label{fig:t-a}
\end{center}
\end{figure}

\section{discussion}
\label{sec:discussion}
We now discuss the details of the results obtained above.
First, we analyze the polaron size by using Gaussian fitting, and describe the overall picture of the transitions as the delocalization of polarons induced by the pressure.
Next, we discuss the transition between site- and bond-centered VBPs and the pinning frequencies of a CDW in terms of Ginzburg-Landau theory.

\subsection{Size of polaron}
\label{ssec:size}
\begin{figure}[htbp]
\begin{center}
\includegraphics[width = 0.9\linewidth]{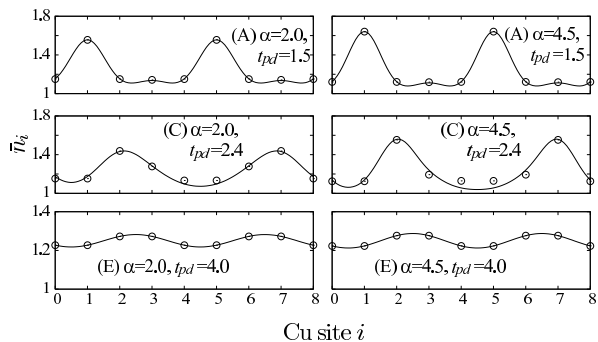}
\caption{hole-density profile fitted by Gaussian.
The left panels for $\alpha=2.0$, and the right for $\alpha=4.5$.
The upper, middle, and lower panels for (A), (C), and (E) states, respectively.
Circles show the hole-density averaged over nearest O sites using the data of the the Hartree-Fock calculation.
Here the profile along only one of the ladder legs is shown.
}\label{fig:profile}
\end{center}
\end{figure}
Fig.~\ref{fig:profile} shows hole-density profiles for several cases.
The data shown by circles in Fig.~\ref{fig:profile} includes averages over nearest O sites:
\begin{eqnarray}
\bar{n}_i
= n_{\mathrm{Cu},i}
\,+ \dfrac{1}{2}\dsum{j
\scriptstyle \mathrm{: nearest}
\atop \scriptstyle \,\,\,\,\,\,\,\,\,\mathrm{neighbors}}{}
n_{\mathrm{O},j},
\end{eqnarray}
where $n_{\mathrm{Cu},i}$ and $n_{\mathrm{O},j}$ are hole densities, respectively, at Cu and O sites obtained by Hartree-Fock calculation.
We fit the data by a least-squares method with the following function:
\begin{eqnarray}
\rho(x,y)
= \dsum{i}{} \dfrac{1}{2\pi\s^2}\, \exp{-\dfrac{(x-\tilde{x}_i)^2
+(y-\tilde{y}_i)^2}{2\s^2}} + h.\nonumber\\
\label{eq:fit}
\end{eqnarray}
Here, the center of the $i$-th polaron is represented by $(\tilde{x}_i,\tilde{y}_i)$, and $h$ takes a value of about 1.1.
In the SP and VBP cases (upper and lower in Fig.~\ref{fig:profile}), the data is well-fitted by the function in Eq.~(\ref{eq:fit}).
In the DBP case, on the other hand, it is not as well-fitted, especially for large $\alpha$ (middle right in Fig.~\ref{fig:profile}).
\begin{figure}[htbp]
\begin{center}
\includegraphics[width = 0.7\linewidth]{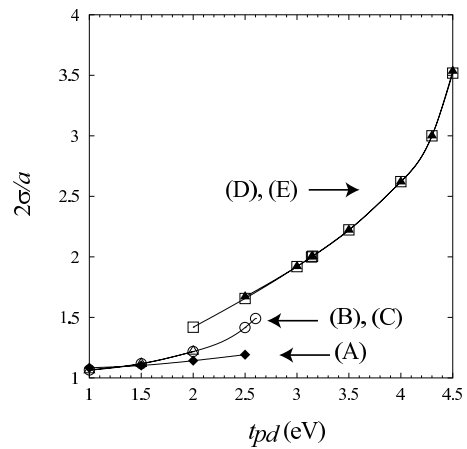}
\caption{$t_{pd}$ dependence of polaron size at $\alpha=2.0$.
(A) filled diamonds, (B) open triangles, (C) open circles, (D) open squares (E) filled triangles.
(B) and (C), and (D) and (E) lie almost on the same line.
}\label{fig:size}
\end{center}
\end{figure}
We define the size of a polaron as $2\s/a$.
The $t_{pd}$ dependence of this polaron size is shown in Fig.~\ref{fig:size}.
In the (A) case, we find $2\s/a$ is between 1.1 and 1.2 within the $t_{pd}$ range between 1.0 and 2.5.
There is little change with increasing $t_{pd}$ in this case.
The (B) state shows the same behavior as the (C) state.
The polaron size of the DBP states increases as $t_{pd}$ increases.
This polaron size growth in the DBP states is greater than in the SP state.
However, the polaron sizes in the VBP are larger than those of SP or DBP.

Comparing with Fig.~\ref{fig:pht-e}, we find that the phase transitions occur from a state of smaller polarons to another state of larger ones.
Especially, we can understand the aspects of the phase transitions (D)$\rightarrow$(E)$\rightarrow$(F) by considering the pinning potential.
Delocalization of the polaron induced by changing $t_{pd}$ causes less distortion of the lattice, as shown in Fig. \ref{fig:pht-gs}.
Therefore, the potential energies for site- and bond-centered states should change:  For small $t_{pd}$, the pinning at Cu is stronger than that at O.
As is discussed in Appendix \ref{app:pinning}, if the change of $t_{pd}$ varies the ratio of the pinning potential at Cu and O sites, then phase transitions are induced.
The phase transitions are also well-described by Ginzburg-Landau theory.
This is discussed in the following section, where, within the Ginzburg-Landau picture, we explain the behavior of the pinning frequencies.

\subsection{Pinning frequency}
\label{ssec:pinning}
We now discuss the transition between site- and bond-centered VBPs and the behavior of the pinning frequencies in terms of a Ginzburg-Landau theory.
We can describe the aspects of the transitions between VBP and UNI, and between site- and bond-centered VBPs by introducing the Landau function (See Appendix \ref{app:GL} for details):
\begin{eqnarray}
F= \int dx\, f[\psi(x),m(x)].
\end{eqnarray}
Here $\psi(x)$ and $m(x)$ are the charge and spin order parameters.
$m(x)$ is the staggered spin density, and $\psi(x)$ is defined by the deviation of the charge density from the uniform state:
\begin{eqnarray}
m(x)&=&(-1)^\frac{x}{a} S(x)\\
\psi(x) &=& \rho(x) - \rho_0.
\end{eqnarray}
In general, $f$ can be written in the following form for an $L$-site-periodic commensurate CDW~\cite{McMillan}:
\begin{eqnarray}
f[\psi(x),m(x)] &=& f_0[\psi(x),m(x)] \nonumber\\
&&+ p(x)\psi(x)^L + q(x)\psi(x)^{2L}.
\end{eqnarray}
Here, $f_0$ is concerned with the lattice-independent spin and charge ordering, and the remaining terms lead to the lattice pinning effect of the CDW.
We assume $f_0$ is of the form
\begin{eqnarray}
f_0[\psi(x),m(x)] &=& r_0 m(x)^2 + u_0 m(x)^4\nonumber\\
&&{}+s_0 m(x)^2\psi(x)+v_0 \psi(x)^2
\end{eqnarray}
with $v_0>0$, that is, the charge order is induced by the magnetic order~\cite{pryadko}.
If we write the charge order parameter in sinusoidal form, the amplitude $\rho_1$ is found to be (see Appendix \ref{app:GL})
\begin{eqnarray}
\rho_1 \propto t_0-t_{pd},
\end{eqnarray}
where $t_0$ is the BP-UNI transition point.

\begin{figure}[ht]
\begin{center}
\includegraphics[width = 0.85\linewidth]{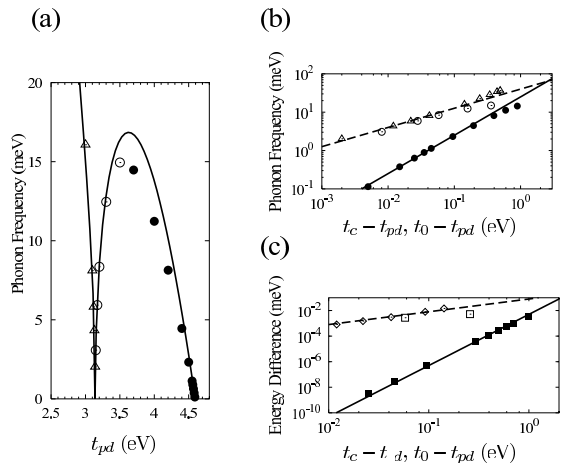}
\caption{Comparison of the functions obtained from Ginzburg-Landau theory with the results from Hartree-Fock and RPA calculations for the $\frac{1}{4}$-hole doping ($L=4$) case.
Lines show the functions (the explicit forms are shown below).
The frequencies from the RPA calculation are plotted with open triangles for $t_{pd} \ltsims t_c$, open circles for $t_{pd} \gtsims t_c$, and filled circles for $t_{pd} \ltsims t_0$.
The energy difference $|E_{(D)}-E_{(E)}|$ from the Hartree-Fock calculation are plotted with open diamonds for $t_{pd} \ltsims t_c$, open squares for $t_{pd} \gtsims t_c$, and filled squares for $t_{pd} \ltsims t_0$.
(a) Pinning frequencies:  $\Omega \propto |t_{pd}-t_c|^{1/2}|t_0-t_{pd}|^{(L-2)/2}$.
(b) Pinning frequencies (log-log plot):  $\Omega \propto (t_0-t_{pd})^{(L-2)/2}$ for $t_{pd} \ltsims t_0$ (solid line) and $\Omega \propto (t_c-t_{pd})^{1/2}$ for $t_{pd} \sim t_c$ (dashed line).
(c) Energy difference between site- and bond-centered VBPs:  $|\Delta F| \propto (t_0-t_{pd})^L$ for $t_{pd} \ltsims t_0$ (solid lines) and $|\Delta F| \propto (t_c-t_{pd})$ for $t_{pd} \sim t_c$ (dashed lines).
}\label{fig:omega4}
\end{center}
\end{figure}
Considering the small oscillations around the equilibrium state,
the pinning frequencies are derived in Appendix \ref{app:GL} as
\begin{eqnarray}
\Omega \propto \left\{
\begin{array}{ll}
|t_{pd}-t_c|^{1/2} & \mbox{for $t_{pd}\sim t_c$}\\
|t_0-t_{pd}|^{(L-2)/2} & \mbox{for $t_{pd}\ltsims t_0$}
\end{array}
\right.,
\label{eq:Omega}
\end{eqnarray}
where $t_c$ is the transition point between site- and bond-centered VBPs.
Both expressions are plotted in Fig.~\ref{fig:omega4} (b), and show a good agreement with the RPA data.
If $t_c$ is not very far from $t_0$, the following form well-describes the behavior of the phonon frequencies as a function of $t_{pd}$ over the whole region around the transition points [see Fig.~\ref{fig:omega4} (a)]:
\begin{eqnarray}
\Omega \propto |t_{pd}-t_c|^{1/2}|t_0-t_{pd}|^{(L-2)/2}.
\label{eq:Omega2}
\end{eqnarray}

As shown in Appendix \ref{app:GL}, the $t_{pd}$ dependence of the energy difference between site- and bond-centered VBPs is given by
\begin{eqnarray}
|\Delta F| \propto  \left\{
\begin{array}{ll}
|t_{pd}-t_c| & \mbox{for $t_{pd}\sim t_c$}\\
|t_0-t_{pd}|^L & \mbox{for $t_{pd}\sim t_0$}
\end{array}
\right..
\end{eqnarray}
These functions are plotted in Fig.~\ref{fig:omega4} (c), where $|E_{(D)}-E_{(E)}|$ from the Hartree-Fock calculation are also plotted for comparison.

\begin{figure}[htbp]
\begin{center}
\includegraphics[width = 0.85\linewidth]{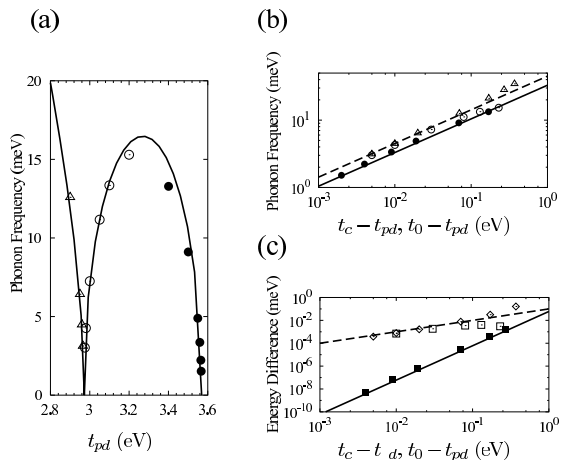}
\caption{Pinning frequencies and energy difference for the $\frac{1}{3}$-hole doping ($L=3$) case.
See the figure caption in Fig.~\ref{fig:omega4}.
}\label{fig:omega3}
\end{center}
\end{figure}

\begin{figure}[htbp]
\begin{center}
\includegraphics[width = 0.85\linewidth]{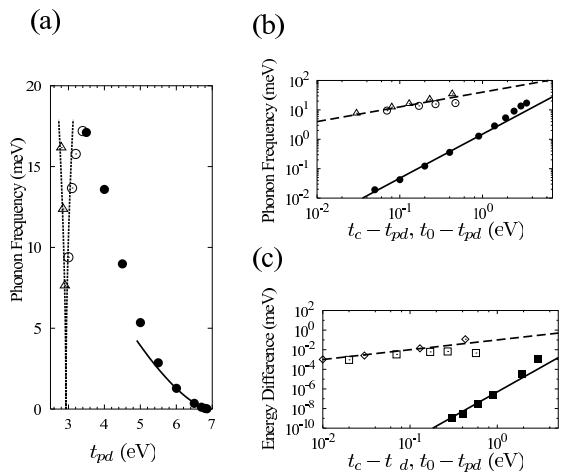}
\caption{Pinning frequencies and energy difference for the $\frac{1}{5}$-hole doping ($L=5$) case.
(a) $\Omega=C(t_0-t_{pd})^{(L-2)/2}$ for $t_{pd}$ around $t_0$ (solid line), $\Omega=C|t_c-t_{pd}|^{1/2}$ for $t_{pd}$ around $t_c$ (dashed line), and the RPA data of the pinning frequencies (same as in Fig.~\ref{fig:omega4}).
(b) and (c) are the same as in Fig.~\ref{fig:omega4}.
}\label{fig:omega5}
\end{center}
\end{figure}

Eq.~(\ref{eq:Omega}) well reproduces the features of the pinned CDW in the $L=4$ case.
Here we investigate energies and the pinning frequencies for different dopings (resulting in different-period CDWs), and further show the validity of Eq.~(\ref{eq:Omega}) for other $L$.
Figs.~\ref{fig:omega3} and \ref{fig:omega5} show the results for the $\frac{1}{3}$-hole doping ($L=3$), and the $\frac{1}{5}$-hole doping ($L=5$) cases.
In the both cases, the results suggest that Eq.~(\ref{eq:Omega}) agrees with the RPA calculation:
The Hartree-Fock and RPA calculations were performed in the systems whose sizes were $6\times2$ ($L=3$) and $10\times2$ ($L=5$).
The $t_{pd}$ dependence of the energy difference between site- and bond-centered VBPs is also well-described by the function obtained from Ginzburg-Landau theory in both cases.

Similar to the $\frac{1}{4}$-hole doping, for $\frac{1}{3}$-hole doping.
Eq.~(\ref{eq:Omega2}) is also a good approximation.
For $\frac{1}{5}$-hole doping, Eq.~(\ref{eq:Omega2}) does not give a good agreement with RPA data.
This is not surprising, since $t_c$ is far from $t_0$ in this case.

\section{summary}
In summary, we have modeled a pressure effect in a Cu$_2$O$_3$ ladder system by using a multi-band Peierls-Hubbard model and simulating the effect of pressure through the hopping strength $t_{pd}$.
With increasing $t_{pd}$, we find a sequence of transitions from SP/BP charge localization to sliding to delocalization, all occurring within a magnetically ordered background.
The ground state has the same number of SPs as doped holes in the case of small $t_{pd}$ ($t_{pd}<1.6$ with the parameters used here).
The ground state configuration changes as SP $\rightarrow$ BP $\rightarrow$ UNI states, as $t_{pd}$ increases.
While SPs are localized and isolated, BPs are partially delocalized.
This means the pressure produces a more delocalized ground state.
In the BP phase, there is also a phase transition between DBP and VBP states.
A similar transition has been found in some other cuprates:
namely, the transition between diagonal- and vertical-stripe states induced by hole doping.

In the VBP phase, we also find a soft mode transition between site- and bond-centered VBP states, although the energies are very close.
Calculations of the phonon eigen-frequency and electronic excitation in the VBP phase yields a sliding mode of VBPs with weak pinning.
The pinning energy in the bond-centered VBP phase is around 15meV at most for $t_{pd}\sim3.5$ eV.
Increasing $t_{pd}$ up to $t_{pd}\sim4.6$ eV makes the pinning zero, and a transition from the VBP to UNI state occurs.
These results suggest experimentally exploring pressure dependence of the low-energy modes found by Kitano \textit{et al.}~\cite{kitano1,kitano2} and Blumberg \textit{et al.}~\cite{blumberg}
If these modes correspond to those we have identified, then their pressure-dependence will follow the interesting pattern in Fig.~\ref{fig:pht-w}.
Studying IR, Raman and optical signatures would further clarify the mode assignments.
The resonant soft X-ray scattering technique of Refs.~\cite{abbamonte,rusydi} could also be used to probe our predicted charge ordering structures as a function of pressure.

The sequence of phases (SP-BP-UNI) is reminiscent of the insulator-metal transition with doping observed in other low-dimensional broken-symmetry groundstate materials, including conjugated polymers~\cite{heeger} and layered cuprates.
It is tempting to associate the mode softening with the onset of a sliding CDW in the spirit of Fr\"{o}hlich.
However, as the phonon-fluctuations soften, additional degrees of freedom (quantum lattice and spin fluctuations) become relevant and need to be considered --- in particular to identify the superconductivity mechanism.
The superconductivity is observed experimentally in a finite range of pressure.
Whether this can be associated with the finite range of $t_{pd}$ with low pinning frequencies (Fig.~\ref{fig:pht-w}) requires comparison with more detailed experiments, but is a tempting scenario.
We have also explored other values of $\alpha$ and found the same general phase sequence as a function of $t_{pd}$ shown in Fig.~\ref{fig:t-a}.

In this study, we mainly considered only the $\frac{1}{4}$-hole-doped case and related commensurate dopings.
Other cases including incommensurate fillings with discommensurations will be reported elsewhere.
We can expect related transitions with doping as with $t_{pd}$, since they should both be controlled by SP or BP wave-function overlaps.

\section*{ACKNOWLEDGEMENT}
This work was supported by the U.S. DOE.

\appendix
\section{Pinning potential}
\label{app:pinning}
Here we attempt to describe the pinning potential of VBP states, and show the details of the discussion regarding the transition between site- and bond-centered VBPs in Sec.~\ref{ssec:size}.
Since we are interested only in VBP states with different phases here, the system we consider can be reduced to one dimension.
Therefore, the pinning potential is a function of $x$, and has a minimum at $x=na$ for the site-centered VBP and at $x=(n+\frac{1}{2})a$ for the bond-centered VBP.
Approximately, the pinning potential may be attributed to the potential at Cu and O sites:
\begin{eqnarray}
E(x)=\dsum{i}{} \big( \mathcal{E}_{\mathrm{Cu},i}(x)
+ \mathcal{E}_{\mathrm{O},i}(x) \big),
\label{eq:E}
\end{eqnarray}
where the phase is chosen such that the state becomes site-centered when $x=0$.
We continue our discussion with the following two assumptions:
(1) both $\mathcal{E}_{\mathrm{Cu}}$ and $\mathcal{E}_{\mathrm{O}}$ are composed of Gaussians,
\begin{eqnarray}
&&\mathcal{E}_{\mathrm{Cu},i}(x)
= -\dfrac{C_1}{\sqrt{2\pi}\s_1}\, \exp{-\dfrac{(x-a\,i)^2}{2\s_1^2}}\\
&&\mathcal{E}_{\mathrm{O},i}(x)
= -\dfrac{C_2}{\sqrt{2\pi}\s_2}\, \exp{-\dfrac{(x-a\,i+\frac{a}{2})^2}{2\s_2^2}},
\end{eqnarray}
and (2) both $\s_1$ and $\s_2$ are comparable to $a$.
The case we consider here can satisfy these conditions.
Since the density profile of VBP states is well-fitted by Gaussians (shown in Sec. \ref{sec:discussion}), we expect the contribution of the \textit{partial} free energy of the Cu and O sites to the \textit{total} is also formed of Gaussians, and that the size of the Gaussians should also be similar to the VBP size.

\begin{figure}[h]
\begin{center}
\includegraphics[width = 0.9\linewidth]{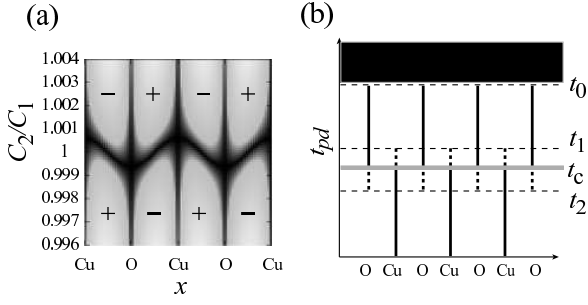}
\caption{(a) The derivative of $E(x)$.
$-$ and $+$ show sign of $dE(x)/dx$.
Blackened area represents $dE(x)/dx\sim0$.
Here both $\s_1$ and $\s_2$ are taken as 0.4.
If we take larger $\s_1$ and $\s_2$, the zig-zag boundary becomes flatter.
(b) Scheme of transitions in bi-polaronic states.
Solid lines show the center of bi-polarons at each $t_{pd}$.
In the case $t_{pd}<t_2$ ($t_1<t_{pd}<t_0$), the system shows the site-centered (bond-centered) VBP ground state.
For $t_2<t_{pd}<t_1$, the pinning potential has two minimal points, and at $t_c$ there is coexistence of site- and bond-centered VBP states.
In $t_0<t_{pd}$ (blackened zone), the (F) state becomes the ground state.
The larger $\s_1$ and $\s_2$, the narrower the double-minimum zone ($t_1<t_{pd}<t_2$) becomes.
}\label{fig:bibun}
\end{center}
\end{figure}

We investigate the potential energy in Eq.~(\ref{eq:E}) with assumption (1).
The pinning potential can be expanded as a Fourier series:
\begin{eqnarray}
E(x)=\dsum{n}{} E_n \cos(n G x),
\end{eqnarray}
where $G=\dfrac{2\pi}{a}$.
The Fourier coefficients are given by the following form:
\begin{widetext}
\begin{eqnarray}
E_n&=& -\dfrac{1}{a}\int_{-\frac{a}{2}}^{\frac{a}{2}}dx
\dsum{i}{} \[ \dfrac{C_1}{\sqrt{2\pi}\s_1}
\,\exp{-\dfrac{(x-a\,i)^2}{2\s_1^2}}
+ \dfrac{C_2}{\sqrt{2\pi}\s_2}
\,\exp{ -\dfrac{(x-a\,i+\frac{a}{2})^2}{2\s_2^2}} \]
\cos{(n G x)}\\
\nonumber\\
&=&-\dfrac{1}{a}
\[ \dfrac{C_1}{\sqrt{2\pi}\s_1}
\,\int_{-\infty}^{\infty}dx\,
\exp{-\dfrac{x^2}{2\s_1^2}}\cos{(n G x)}
+(-1)^n\,\dfrac{C_2}{\sqrt{2\pi}\s_2}
\,\int_{-\infty}^{\infty}dx\,
\exp{-\dfrac{x^2}{2\s_2^2}}\cos{(n G x)}
\]\\
\nonumber\\
&=& -\[ \dfrac{{C}_1}{a} \,\exp{-\dfrac{n^2 G^2 \s_1^2}{2}}
+(-1)^n\,\dfrac{{C}_2}{a} \,\exp{-\dfrac{n^2 G^2 \s_2^2}{2}}\].
\label{eq:four}
\end{eqnarray}
\end{widetext}
From Eq.~(\ref{eq:four}), it follows that $E_n$ for large $n$ vanishes.
By considering the fact that the Gaussian almost vanishes at three-fold half maximum full-width, we evaluate the condition to neglect the components as:
\begin{eqnarray}
\dfrac{n^2 G^2 \s^2}{2} = \dfrac{2\pi^2 n^2 \s^2}{a^2}
\gg \dfrac{3^2}{2}.
\end{eqnarray}
Here all variables are positive, so that the condition is given by
\begin{eqnarray}
\dfrac{2\s}{a}&\gg&\dfrac{1}{n}.
\end{eqnarray}
If we consider the case that both $\s_1$ and $\s_2$ are comparable to $a$ (assumption (2)), $E(x)$ is approximately represented by the cosine curve or a slightly modified one:
\begin{eqnarray}
E(x)\sim E_0 + E_1 \cos{(G x)}+ E_2 \cos{(2G x)}.
\end{eqnarray}
Here $E_2$ is small and it does not change the shape of $\cos{(G x)}$ very much unless $E_1$ is small as well as $E_2$.
In such a situation, only one of the site- or bond-centered VBP states is stable, and is determined by the relation between the $C_1$ and $C_2$ magnitudes.
However, if $E_1$ is smaller than $4E_2$, $E(x)$ has minima at both Cu and O sites.
This can happen when $C_1 \sim C_2$ (see Fig.~\ref{fig:bibun}~(a)).

In Fig.~\ref{fig:bibun}~(b), the scheme of phase transition which follows from the Hartree-Fock calculation (Fig.~\ref{fig:pht-e}) is shown.
From the phonon frequency calculation (Fig.~\ref{fig:pht-w}), the bi-stability interval, $t_1-t_2$, is very narrow and difficult to identify numerically.
Comparing this scheme and the behavior of $dE(x)/dx$, we understand the phase transition (D)$\rightarrow$(E)$\rightarrow$(F) as follows.
(1) Since the site-centered state (D) is found at small $t_{pd}$ ($<t_2$), $C_2/C_1$ has to be less than 1 in this case.
(2) Since the double-minimum region is very narrow ($t_1 \sim t_2$), the zig-zag boundary near $C_2/C_1\sim1$ in Fig.~\ref{fig:bibun} is almost flat when $t_{pd}$ is close to the transition point.
(3) Since the bond-centered state (E) is found above the transition point ($>t_1$), $C_2/C_1$ should be found in the upper region ($>1$) across the coexistence point $t_c$.
(4) Increasing $t_{pd}$ far above $t_1$, $E(x)$ becomes close to a constant $F_0$, which implies that the transition from (E) to (F) occurs.

\section{Ginzburg-Landau theory of the site-centered to bond-centered VBP transition}
\label{app:GL}
Here we give details of the discussed transition between site- and bond-centered VBPs in Sec. \ref{ssec:pinning}.
First, we describe the statics of the transition by considering the free energy.
Then, by considering small long-wavelength oscillations around the ground state, we estimate the pinning frequencies.

We start with the Landau function:
\begin{eqnarray}
F= \int dx\, &\big[& r_0 m(x)^2 + u_0 m(x)^4\nonumber\\
&&{}+s_0 m(x)^2\psi(x)+v_0 \psi(x)^2\nonumber\\
&&{}+ p(x)\psi(x)^L + q(x)\psi(x)^{2L} \, \big]
\end{eqnarray}
We write $p(x)$ and $q(x)$ in an expanded form, for example
\begin{eqnarray}
p(x)= \sum_{n=1}^{\infty} p_n \cos(nGx).
\end{eqnarray}
As seen in Sec. \ref{sec:discussion}, the density profile of the VBP is well fitted by large-size Gaussians, so we can neglect higher harmonic terms of the order parameters and describe them by sinusoidal waves:
\begin{eqnarray}
&&\mbox{$\psi(x,\phi) \sim \rho_1 \cos(\frac{G}{L} x + \frac{\phi}{L})$},\\
&&\mbox{$m(x,\phi) \sim m_1 \sin(\frac{G}{2L} x + \frac{\phi}{2L})$}.
\end{eqnarray}
We may constrain $\rho_1\ge0$.

We evaluate the free energy using the following integral ($l\neq0$):
\begin{eqnarray}
&&\hspace{-1cm}
\int_V\frac{dx}{V}
\mbox{$\cos^{n}(\frac{G}{L}x+\frac{\phi}{L})$}
\cos(lGx)\nonumber\\
&&=\sum_{h=0}^{(n-1-\tilde{\delta}_{n})/2}
\frac{1}{2^{n}}{n \choose h}
\,\delta_{n-2h,lL}
\,\cos(l\phi),
\end{eqnarray}
where ${n \choose l}=\frac{n!}{(n-l)!l!}$, $\delta$ is Kronecker's delta, and $\tilde{\delta}_n$ is unity for even $n$ and zero for odd $n$.
Integrating over volume $V$, the free energy per unit volume is
\begin{eqnarray}
F_\phi(t_{pd},\rho_1,m_1)
&=&F^{(0)}(t_{pd},\rho_1,m_1) \nonumber\\
&&{}+\left[\tilde{p}_1 \rho_1^L
+\tilde{q}_1\rho_1^{2L} \right]
\cos(\phi)\nonumber\\
&&{}+ \tilde{q}_2 \rho_1^{2L} \cos(2\phi),
\label{eq:F_phi}\\
F^{(0)}(t_{pd},\rho_1,m_1)&=&
\tilde{r}_0 m_1^2+\tilde{u}_0m_1^4\nonumber\\
&&{}-\tilde{s}_0m_1^2\rho_1
+\tilde{v}_0\rho_1^2,
\label{eq:f0}
\end{eqnarray}
where
\begin{eqnarray}
&\tilde{p}_1=\frac{1}{2^L}p_1,\hspace{0.3cm}
\tilde{q}_1=\tilde{\delta}_{L}{ 2L \choose \frac{L}{2}}
\frac{1}{2^{2L}} q_1,\hspace{0.3cm}
\tilde{q}_2=\frac{1}{2^{2L}}q_2,&\\
&\tilde{u}_0=\frac{3}{8}u_0,\hspace{0.3cm}
\tilde{v}_0=\frac{1}{2}v_0,\hspace{0.3cm}
\tilde{r}_0=\frac{1}{2}r_0,\hspace{0.3cm}
\tilde{s}_0=\frac{1}{4}s_0,&
\end{eqnarray}
Note that the signs of these variables are the same with or without tilde.
The $r_0$ and $u_0$ terms govern the BP-UNI continuous transition, and the rest characterize the transition between site- and bond-centered VBPs.
Subsequently, we neglect the $\tilde{q}_1\rho_1^{2L}$ term in Eq.(\ref{eq:F_phi}), since this term would be smaller than the $\tilde{p}_1\rho_1^{L}$ term.

First, we consider the BP-UNI transition.
Phenomenologically assuming $r_0\propto t_{pd}-t_0$ ($t_0$ is the BP-UNI transition point, where $p_1=0$) and $u_0>0$, we find $m_1 \propto (t_0-t_{pd})^{1/2}$ for $t_{pd}<t_0$.
From the last two terms of Eq.~(\ref{eq:f0}), it follows that
\begin{eqnarray}
\rho_1 \propto m_1^2 \propto t_0-t_{pd}
\end{eqnarray}
for $t_{pd}<t_0$.
Next, we consider the transition between site- and bond-centered VBPs.
In the case $t_{pd}\neq t_c$ ($t_c$ is the coexistence point), the minimum point of this free energy is controlled by the $p_1$ term.
At $t_{pd}=t_c$, the $p_1$ term vanishes, and the minimum of the free energy is determined by the $q_2$ term.

The minimum point is given by
\begin{eqnarray}
\phi=\left\{
\begin{array}{rl}
\pi & \mbox{for $p_1>0$}\\
 0  & \mbox{for $p_1<0$}
\end{array}
\right..
\end{eqnarray}
$\phi=\pi$ corresponds to the bond-centered case at $t_{pd}>t_c$, and $\phi=0$ to the site-centered case at $t_{pd}<t_c$.
At the critical point ($t_{pd}=t_c$), the $p_1$ term vanishes; and for
\begin{eqnarray}
q_2\rho_1^{2L}<0
\end{eqnarray}
there are minima at both $\phi=0$ and $\pi$.

Next we consider the small oscillations around the equilibrium state:
$\phi=\phi_0+\delta\phi$ ($\phi_0$ takes either 0 or $\pi$ for the ground state).
Using $p_1\cos(\phi_0)=-|p_1|$ and $\cos(2\phi_0)=1$, the free energy is expanded for $\delta\phi$ as
\begin{eqnarray}
&&\hspace{-1cm}F_{\phi_0+\delta\phi}(t_{pd},\psi,m)
\nonumber\\
&=& F^{0}(t_{pd},\rho_1,m_1)
+ \tilde{p}_1 \rho_1^L \cos(\phi_0)
\left[1-\frac{(\delta\phi)^2}{2}\right]\nonumber\\
&&{}+ \tilde{q}_2 \rho_1^{2L}
 \cos(2\phi_0)\left[1-\frac{(2\delta\phi)^2}{2}\right]\\
&=& \left[F^{0}(t_{pd},\rho_1,m_1)
- |\tilde{p}_1| \rho_1^L
+ \tilde{q}_2 \rho_1^{2L}\right]\nonumber\\
&&{}+\frac{1}{2}\left( |\tilde{p}_1| \rho_1^L
- 4 \tilde{q}_2 \rho_1^{2L}
\right)(\delta\phi)^2.
\label{eq:deltaF}
\end{eqnarray}
The second line of Eq.~(\ref{eq:deltaF}) is used to find mode frequencies.
The Lagrangian for the oscillation of VBPs is given by
\begin{eqnarray}
\mathcal{L}=\frac{M}{2}\frac{d(\delta\phi)}{dt}
-\frac{M\Omega^2}{2}(\delta\phi)^2
+\mathrm{const}.
\end{eqnarray}
Here, $M$ is the effective mass of the CDW, and
\begin{eqnarray}
\Omega^2
=\frac{1}{M}
\left(|\tilde{p}_1| \rho_1^L
- 4 \tilde{q}_2 \rho_1^{2L}
\right).
\end{eqnarray}
Using $p_1 \propto t_{pd}-t_c$ and $\rho_1 \propto t_0-t_{pd}$, and supposing $M\propto\rho_1^2$, for $t_{pd}<t_0$,
we find
\begin{eqnarray}
\Omega \propto \left\{
\begin{array}{ll}
|t_{pd}-t_c|^{1/2} & \mbox{for $t_{pd}\sim t_c$}\\
|t_0-t_{pd}|^{(L-2)/2} & \mbox{for $t_{pd}\ltsims t_0$}
\end{array}
\right..
\end{eqnarray}

It also follows from Eq.~(\ref{eq:F_phi}) that the $t_{pd}$ dependence of the energy difference between site- and bond-centered VBPs, $|\Delta F|$, is given by
\begin{eqnarray}
|F_{\phi=0}-F_{\phi=\pi}| \propto  \left\{
\begin{array}{ll}
|t_{pd}-t_c| & \mbox{for $t_{pd}\sim t_c$}\\
|t_0-t_{pd}|^L & \mbox{for $t_{pd}\sim t_0$}
\end{array}
\right..
\end{eqnarray}

\end{document}